\documentclass[aps,prappl,twocolumn,superscriptaddress,notitlepage]{revtex4-2}
\usepackage{CJK}
\usepackage[pdftex]{graphicx}
\usepackage{float}
\usepackage{amsmath}
\usepackage{amssymb}
\usepackage{url}
\usepackage{natbib}
 
\usepackage[colorlinks,
                   linkcolor=blue,
                   anchorcolor=blue,
                   citecolor=blue,
                   urlcolor=blue]
                  {hyperref}              
\begin{document}
\begin{CJK*}{UTF8}{gbsn}
\title{Self-driven Hybrid Atomic Spin Oscillator}
\author{Erwei Li}
\thanks{These authors contributed equally to this work and should be regarded as co-first authors.}
\author{Qianjin Ma}
\thanks{These authors contributed equally to this work and should be regarded as co-first authors.}
\author{Guobin Liu}
\email[Corresponding author:\,] {liuguobin@ntsc.ac.cn}
\author{Peter Yun}
\author{Shougang Zhang}
\affiliation{National Time Service Center, Chinese Academy of Sciences, Xi'an, 710600, China.}
\affiliation{University of Chinese Academy of Sciences, Beijing, 100049, China}

\date{\today}
\begin{abstract}
A self-driven hybrid atomic spin oscillator is demonstrated in theory and experiment with a vapor Rb-Xe dual-spin system. The raw signal of Rb spin oscillation is amplified, phase-shifted and sent back to drive the Xe spins coherently. By fine tuning the driving field strength and phase, a self-sustaining spin oscillation signal with zero frequency shift is obtained. The effective coherence time is infinitely prolonged beyond the intrinsic coherence time of Xe spins, forming a hybrid atomic spin oscillator. Spectral analysis indicates that a frequency resolution of 13.1 nHz is achieved, enhancing the detection sensitivity for magnetic field. Allan deviation analysis shows that the spin oscillator can operate in continuous wave mode like a spin maser. The prototype spin oscillator can be easily implanted into other hybrid spin systems and enhance the detection sensitivity of alkali metal-noble gas comagnetometers.
\end{abstract}
\maketitle
\end{CJK*}

Alkali metal-noble gas comagnetometer has been used for both fundamental and practical applications, such as the search for axion like particles or new physics beyond the standard model \cite{Bulatowicz2013PRL, Domainwall2013PRL, Limes2018PRL, EDM2019PRL} and inertial navigation gyroscope\cite{Donley2010NMRG, Kornack2005PRL, Walker2016NMRG}. It is important for the comagnetometer to continuously improve the detection sensitivity to magnetic field or angular velocity, to beat the frequency or energy limits at record-breaking measurements \cite{Limes2018PRL, Frequencyshift2019PRA, Romalis2020PRL}. 

Basically, comagnetometer sensitivity is determined by two main factors, the coherence time and signal to noise ratio (SNR) of the spin oscillation signal. To improve the SNR, various parametric modulations together with lock-in detection are regularly utilized to suppress the noise \cite{Budker2002RMP, Limes2018PRL, Walker2019PRA, Romalis2020PRL}. Signals can be amplified using smart optical design such as multi-pass vapor cells \cite{Li2011PRA, Sheng2021PRA}. As for the coherence time, increasing the spin longitudinal relaxation time is the first consideration. Buffer gas filling \cite{buffergas1957PR} and anti-relaxation paraffin coating \cite{coating1966PR} in vacuum atomic vapor cells have been used in atomic clocks and magnetometers for long time, so the techniques are inherited naturally by alkali metal-noble gas comagnetometers. Multiple spatial or temporal spin-field interactions, such as the well known separated oscillating fields (producing the well known Ramsey fringe in atom fountain clocks) \cite{Ramsey1949PR}, Hahn-echo or spin-echo and CPMG pulse excitation techniques in nuclear magnetic resonance (NMR) spectroscopy \cite{SpinEcho1950PR, CPMG1958RSI, spindynamics} can also effectively increase the transverse relaxation time or coherence time of spin oscillations.

Inspired by the self-oscillating nonlinear magneto-optical rotation (NMOR) magnetometer by Kitching et al. \cite{Kitching2005RSI} and following the feedback oscillator electronics and ac magnetic field excitation NMR, we have been long wondering whether a self-sustaining spin oscillator based on alkali metal-noble gas comagnetometer is possible. Compared to the magnetometer using pure alkali metal atoms, an alkali metal-noble gas comagnetometer has the advantage of much longer nuclear spin relaxation time, typically from tens of seconds to even hours for example with $^3$He \cite{Polarized3He2017RMP}, and the nuclear spin coherence can be transferred to alkali atom spin in the hybrid spin dynamics \cite{Kornack2002PRL}. It is thus theoretically easier for such a system to become self-oscillating given proper feedback conditions. Here we propose a self-driven hybrid atomic spin oscillator theoretically and demonstrate its feasibiliy by a simple-to-establish experimental apparatus.

We consider a model Rb-Xe comagnetometer under the self-feedback or self-driving mode, where the driven spin dynamics can be described by the coupled Bloch equations
\begin{equation}
\begin{split}
\frac{\partial{\bf M}^{\rm Rb}}{\partial t}&=\frac{\gamma_{\rm Rb}}{q}{\bf M^{\rm Rb}}\times({\bf B}_0+ {\lambda} {\bf  M^{\rm Xe}})\\
&\qquad+\frac {M_0^{\rm Rb}{\hat z}-\bf M^{\rm Rb}} {\it{q T^{\rm Rb}}},\\
\frac{\partial{\bf M}^{\rm Xe}}{\partial t}&=\gamma_{\rm Xe}{\bf M^{\rm Xe}}\times({\bf B}_0+\lambda{\bf M^{\rm Rb}}+G M_x^{\rm Rb}e^{i\theta}{\hat y})\\
&\qquad+\frac{M_0^{\rm Xe}{\hat z}-{\bf M^{\rm Xe}}} {{\it T}^{\rm Xe}}.
\label{eq1}
\end{split}
\end{equation}
where $\bold M^{\rm Rb}$ and $\bold M^{\rm Xe}$ are the Rb and Xe spin magnetizations respectively, $M_0^{\rm Rb}\hat z$ and $M_0^{\rm Xe}\hat z$ are the initial Rb and Xe spin magnetizations in $z$ direction respectively. $\bold B_0$ is the static magnetic field, $T^{\rm Rb}$ and $T^{\rm Xe}$ are the Rb and Xe spin relaxation times respectively. $\it q$ is the slowing down factor due to Rb-Xe spin exchange collisions in high temperature. $\lambda$=8$\pi \kappa$/3 ($\kappa\simeq$500 for Rb-Xe spin-exchange interaction) is the enhancement factor due to Fermi contact interaction between Rb valence electrons and Xe nuclei \cite{Kornack2002PRL}. $\gamma_{\rm Rb}$ and $\gamma_{\rm Xe}$ are the gyromagnetic ratios for Rb and Xe spin respectively. Other unspecified notations and the numerical solving procedure can be referred to our previous work in \cite{Liu2019PRA}.

The driving term $\it GM_x^{\rm Rb}e^{i\theta} \hat y$ is the phase-shifted spin oscillation of Rb spins in transverse direction and then it couples with Xe spins transversely like a conventional NMR excitation pulse protocol. In this case, the equivalent spin dynamics of the comagnetometer is: first the fast relaxing Rb spins adiabatically follow the slow precessing Xe spins under the overdamping condition as in classical driven oscillators \cite{Liu2021PRA}; secondly, the Rb spin oscillation signal is amplified, phase shifted and sent back to drive the Xe spins as an ac magnetic field. Thus the hybrid spin system forms a self-driven oscillator.  For experimental realization, the Rb spins also act as a probe, reading out Xe spin precession signal. The driving field strength is determined by the gain factor $G$ and the driving field is in-phase or out of phase depending on the phase shift $\theta$.

The comagnetometer works in two modes: open-loop mode ($\it G$=0) and close-loop mode ($\it G$$\neq$0). In close-loop mode, the self-driving comagnetometer can work in two different states depending on the magnitude of $\it G$. In classical oscillator electronics, a close loop system can be made self-oscillating when product of the gain and feedback approaches one. In simulation, we found a similar critical condition for the close loop comagnetometer to become self-oscillating, which in this case is described as
\begin{equation}
\begin{split}
G M_0^{\rm Rb} \simeq \lambda M_0^{\rm Xe}/q,
\label{eq2}
\end{split}
\end{equation}
meaning the self-oscillating can be triggered when the driving field strength approaches the 1/$q$th of the initial Xe spin magnetization. It is key to note that the signal amplitude of comagnetometer output in open loop is proportional to the Rb spin magnetization strength while the long time oscillation behavior is governed by the Xe spin precession. This makes the Eq.\ref{eq2} taking the present form. The driving mechanism can be understood as a variation to the conventional $\pi$-pulse Hahn echo sequence \cite{SpinEcho1950PR}. The excitation or driving filed is an always near-resonant wave with moderate strength, rather than a short, strong and broad-band radio-frequency pulse. As the driving field circulates continuously in the close loop mode, multiple echos can present and persist to different time lengths depending on $G$. When the driving field reaches a threshold strength that the amplitude of echos ceases to decrease,  the spin oscillation reaches a steady state. 

However, we have to point out that the Eq.\ref{eq2} is an empirical relation concluded by numerical simulation results. There are probably other factors, such as the intrinsic coherence time of nuclear spins $T^{\rm Xe}$, playing a certain role in the driving mechanism. However, we couldn't change $T^{\rm Xe}$ much once the experimental setup is constructed, so here we take the $T_2^{\rm Xe}$ as a fixed parameter ($\sim$10 seconds). For typical experimental conditions, the initial Rb and Xe spin magnetizations are at the same level, i.e. $\lambda M_0^{\rm Rb}\simeq \lambda M_0^{\rm Xe} \sim$0.1 mG \cite{Bulatowicz2013PRL}. So for the $^{87}$Rb-$^{129}$Xe spin pairs, the threshold driving field strength concluded by the simulation results is $G_{\rm sim} M_0^{\rm Rb} \simeq \lambda M_0^{\rm Xe}/q\sim$20 $\mu$G, i.e. $G_{\rm sim}\sim$1000.

\begin{figure}[htbp]
\centering
\includegraphics[width=0.5\textwidth]{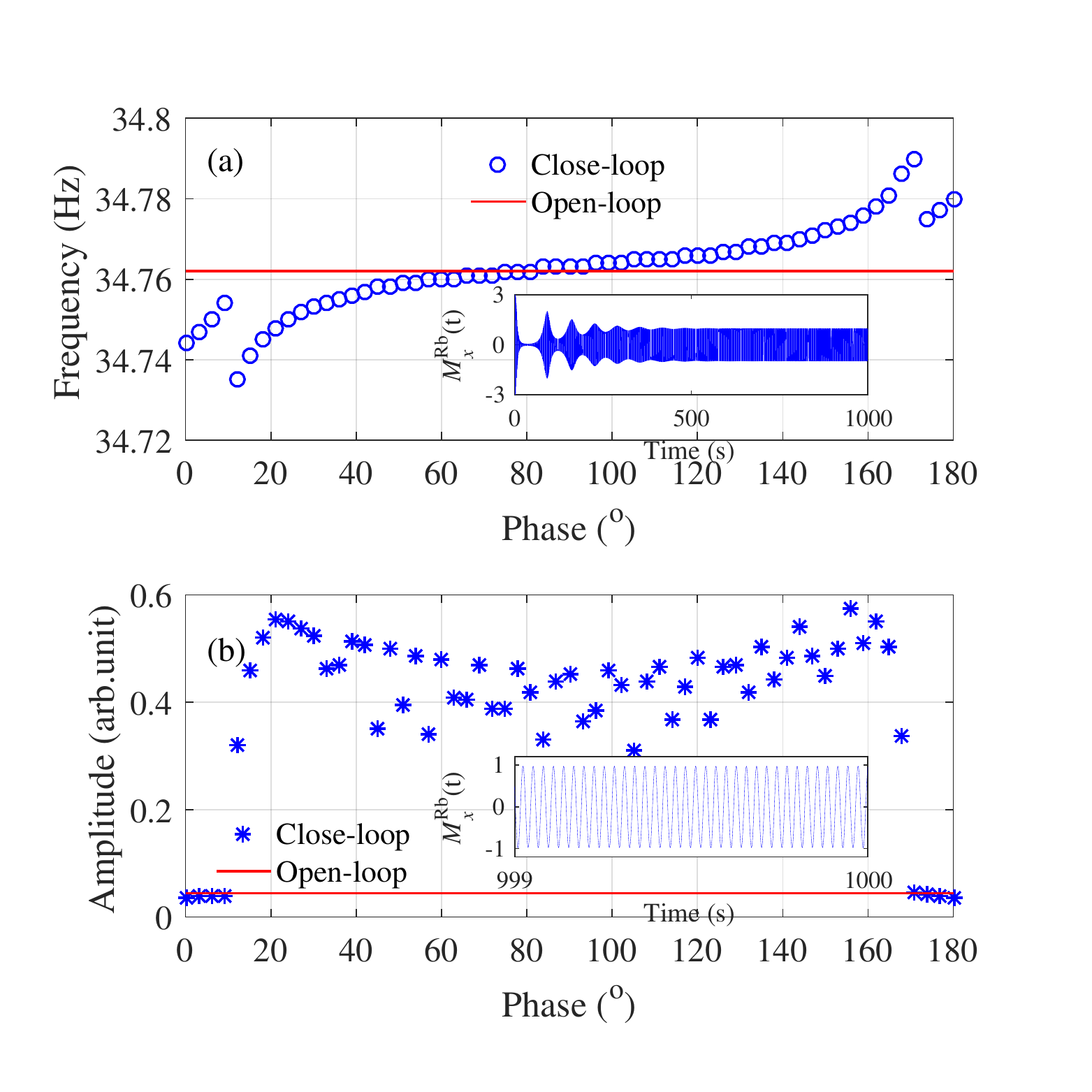}
\caption{(color online) Simulated self-driving spin oscillation signal (a) in close loop mode and its frequency response as a function of the self-driving phase shift $\theta$ at a fixed strong gain $G$=1000 (b). The crossing between the open-loop curve (red line) and the close-loop curve (blue circles) indicate there is a phase point where zero frequency shift (ZFS) occurs albeit the presence of Bloch-Siegert shift effect under strong off-resonance self-driving field.}
\label{fig1}
\end{figure}

Generally, the self-driving field strength can be divided into two regions: 1) When $\it GM_0^{\rm Rb}$$\le$$\lambda M_0^{\rm Xe}/q$, the self-driving field is said to be weak since the Xe spin oscillation decays exponentially as usual; 2) When $\it GM_0^{\rm Rb}$$\ge$$\lambda M_0^{\rm Xe}/q$, the self-driving effect becomes strong and a self-sustaining spin oscillation emerges, with an effective coherence time far longer than the intrinsic spin relaxation time of $^{129}$Xe spins in typical temperature and buffer gas conditions, as shown by the two insets in Fig.\ref{fig1}. 

The phase is also critically important in realization of the self-sustaining spin oscillation. Simulation results show that the self-sustaining oscillation can persist for a phase range determined by
\begin{equation}
\begin{split}
\theta_0<\theta<180^{\circ} - \theta_0,
\label{eq3}
\end{split}
\end{equation}
as indicated by the two sharp structures in the frequency-phase (dispersions) and amplitude-phase (steps) diagrams in Fig.\ref{fig1}. The greater the $G$, the smaller the $\theta_0$, meaning a wider phase range  for self-sustaining oscillation exists given stronger self-driving field.

It is also shown in Fig.\ref{fig1}(b), the frequency shifts almost linearly with $\theta$ in most of the self-sustaining oscillation phase range given by Eq.\ref{eq3} except at around the critical phase points at $\theta_0$ and $180^{\circ}$$-$$\theta_0$. The shift can be explained as follows, due to the well known Bloch-Siegert shift effect in NMR spectra \cite{BSshift}, the mismatch of the initial phase values between the driving field signal and open loop signal can lead to an accumulation effect, which may keep changing gradually the close loop frequency. In this case, one may think the self-driven comagnetometer not suitable for the precision measurement purpose despite the potential advantages in signal amplification and coherence time prolonging. 

However, by numerically solving the Eq.\ref{eq1}, we find a phase point where the spin oscillation frequency shift vanishes. As shown in Fig.\ref{fig1}(a), the frequency of close-loop (blue circles) crosses with that of open-loop (red line), implying that frequency shift vanishes at certain phase value. Theoretically the in-phase driving phase value is 90$^{\circ}$ considering the $y$-axis excitation and $x$-axis detection configuration, however, due to Bloch-Siegert shift effect, the zero frequency shift (ZFS) phase $\theta_{\rm ZFS}$ is about several degrees below 90$^{\circ}$. The position of $\theta_{\rm ZFS}$ depends also on $G$. The larger the $G$, the further the $\theta_{\rm ZFS}$ away from 90$^{\circ}$. The existence of the ZFS phase is important as we have to rule out any possible non-systematic frequency shift sources \cite{Frequencyshift2019PRA} to find out the fundamentally unknown spin-dependent interactions.
\begin{figure}[htbp]
\centering
\includegraphics[width=0.5\textwidth]{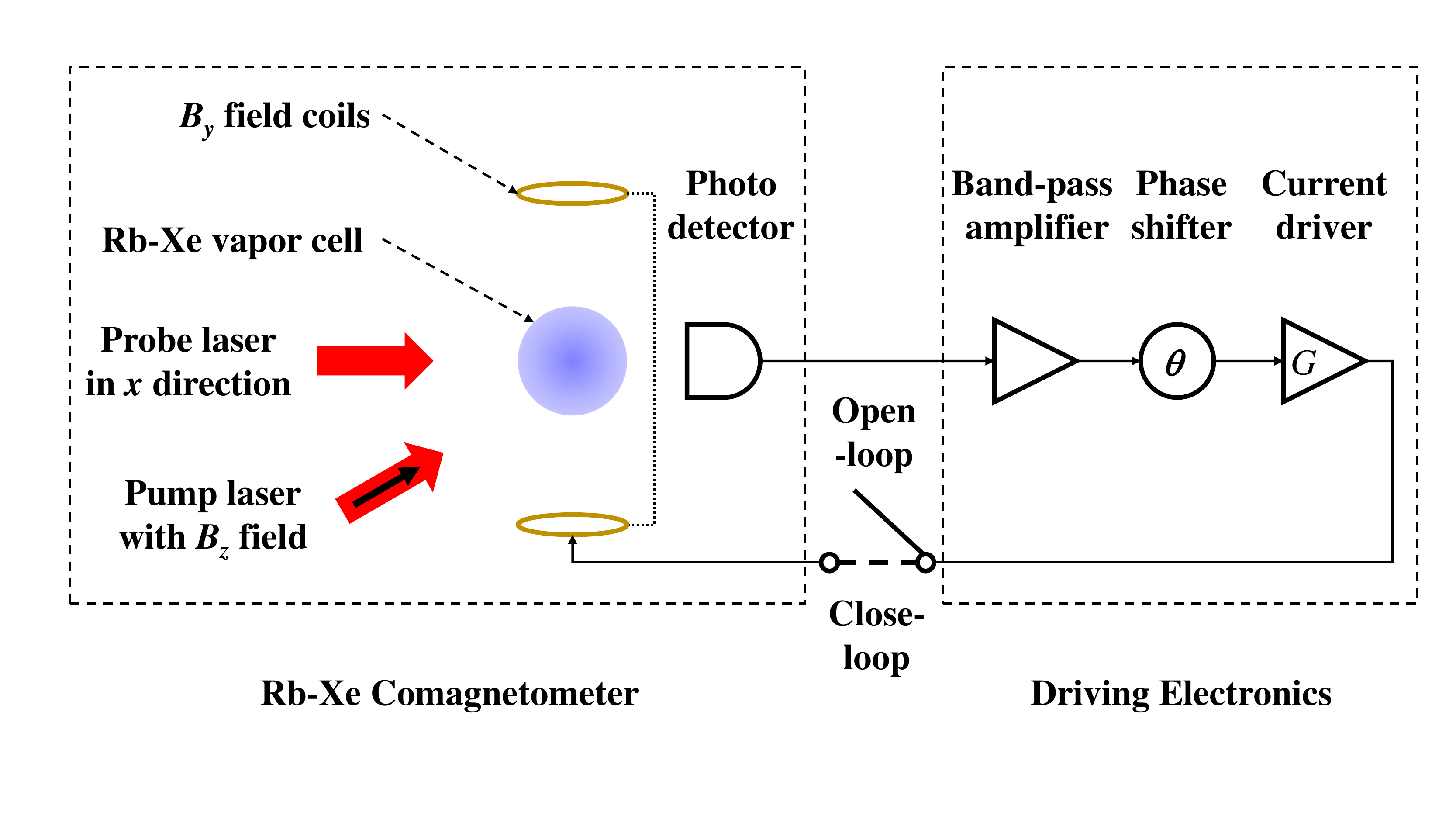}
\caption{Experimental schematic of the self-driving Rb-Xe spin oscillator. It consists of a typical pump-probe Rb-Xe comagnetometer (left) and a driving electronics with tunable gain and phase parameters (right). The pump and probe laser powers are 54 mW and 3 mW, respectively. The static field $B_z$ is $\sim$30 mG, corresponding to a $^{129}$Xe spin oscillation frequency $\nu_0$$\sim$35 Hz.}
\label{fig2}
\end{figure}

To verify above simulation results, we have constructed a Rb-Xe comagnetometer setup and specially designed the driving electronics with tunable gain and phase parameters, as depicted in Fig.\ref{fig2}.    

The main part is a typical Rb-Xe comagnetometer, with a vacuum atomic vapor cell containing Rb-Xe mixture gas at high temperature ($\sim$120 $^{\circ}$C) as the atom spins media. The cell is spherical with a diameter of about 10 mm, containing 4 torr $^{129}$Xe and 35 torr $^{131}$Xe and several tens of torr N$_2$ as buffer gas. The cell in placed in the center of three sets of Helmholtz coils orthogonally with each other. The three sets of coils have current to magnetic field convertion coefficient $C_{\rm B-I}$$\sim$1-2 mG/mA. For the $B_y$ coil, it is 2 mG/mA. The driving current can be changed by fine tuning a rheostat (from 1 to 100 k$\Omega$) in series with the coil. A circularly polarized 795 nm laser with power of ~54 mW along the $z$ direction shines into the cell to align the Rb atom spins. Then the Rb spin polarization was transferred to Xe atoms spins via rapid spin-exchange collisions \cite{Happer1997RMP}. The polarized Xe spins drive Rb spins in a classical way \cite{Liu2021PRA} and at last a linearly polarized 780 nm laser with power of ~3 mW along the $x$ direction reads out the Rb spin dynamics over time as the comagnetometer original output signal. Here for simplicity, we use single symbol for photodetector, while it actually include high extinction ratio linear polarizer and polarized beam spllitter froming a high sensitivity optical polarimeter detection scheme \cite{Budker2002RMP}.
 
The driving electronics are basically a combination of a band-pass amplifier, a phase shifter and a current driver. The center frequency of the band-pass filter is 35 Hz with a $Q$ factor of 10, meaning a passing-band frequency range of $\sim$35$\pm$1.75 Hz. Thus the $^{131}$Xe spin oscillation signal at $\sim$10 Hz is suppressed in the close loop mode. Two factors affect the design of driving electronics. First, the Rb spins have a larger response to Xe spins in lower frequency. Secondly, for typical applications including magnetic field and rotation rate measurements, the Rb-Xe comagnetometer works usually in an ultra-low frequency range. Both factors urge the target spin oscillation to be working at the range from several hertz to a few tens of hertz. Unfortunately, the 1/$f$ law of noise spectra indicates the spin oscillation signal may be easily disturbed by loud amplitude and phase noises at the frequency range. While the gain $G$ has high tuning resolution, the phase resolution $\Delta\theta$ is limited to a few degrees. 

\begin{figure}[htbp]
\centering
\includegraphics[width=0.5\textwidth]{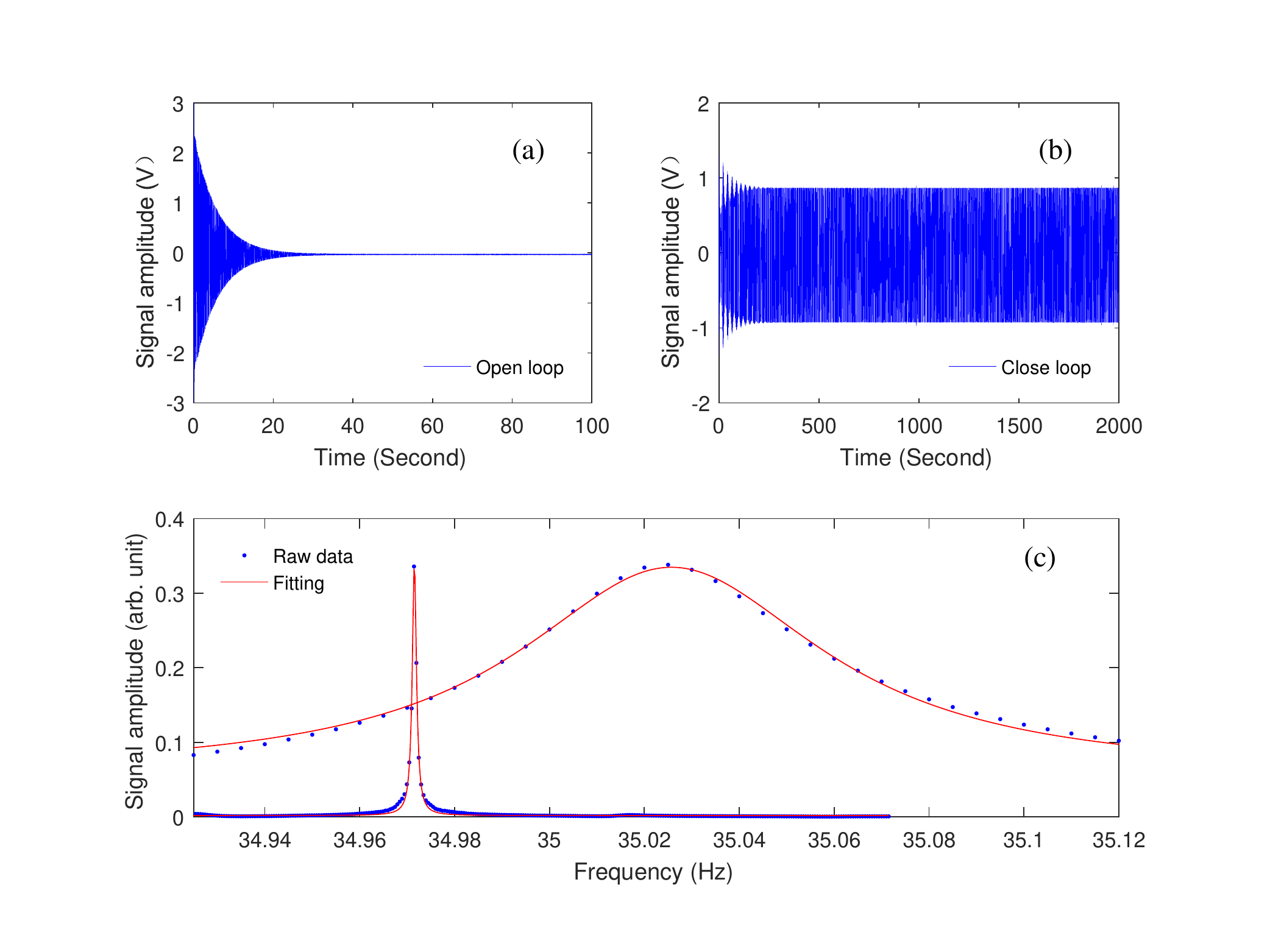}
\caption{(color online) Spin oscillation signals of the self-driving Rb-Xe spin oscillator in open-loop (a) and close-loop (b) modes, and the corresponding Fourier spectra comparison (c). A linewidth narrowing by a factor of 75 and SNR enhancement by a factor of 34 is achieved when switching from the open loop to close loop modes. The open loop Fourier spectra in (c) was amplified by ten times for increasing the visibility. }
\label{fig3}
\end{figure}

For one typical experimental cycle, we first break the link between the Rb-Xe comagnetometer and driving electronics and record an open-loop spin oscillation signal, as shown in Fig.\ref{fig3}(a). Then we restore the link, set the electronics driving output at a fixed gain and change the phase point by point with an accuracy of a few degrees, recording the spin oscillation signals accordingly. At last, we fix the phase to the ZFS point where the close-loop spin oscillation frequency coincides with the open-loop one and record a long-time spin oscillation signal, as shown in Fig.\ref{fig3}(b). 

It was experimentally found that the self-oscillation was triggered when the driving current of the $B_y$ coil reaches $\sim$20$\mu$A, i.e. a driving field strength $B_y$$\sim$40 $\mu$G. By measuring the frequency shift of Xe spin precession $\Delta\nu_{\rm Xe}$ while reversing the pump laser polarization from $\sigma^+$ to $\sigma^-$, the Rb spin magnetization field experienced by Xe spins is determined to be $\Delta\nu_{\rm Xe}$/$\gamma^{\rm Xe}$=0.1 Hz/11.78 MHz/T=0.085 mG in our system, so the Rb spin magnetization field in vacuum is 0.085/$\lambda \sim$0.02 $\mu$G. Therefore the experimental threshold gain factor is $G_{\rm exp}$=$B_y$/$M_0^{\rm Rb}$$\simeq$2000. This threshold driving field strength is about two times larger than the simulated results. Considering the uncertainty in the estimated values of $M_0^{\rm Rb}$ and $M_0^{\rm Xe}$ and other parameters (such as the $\kappa$ and $T_2^{\rm Xe}$), the experimental results agree well with the simulations within the allowed error range. 

\begin{figure}[htbp]
\centering
\includegraphics[width=0.5\textwidth]{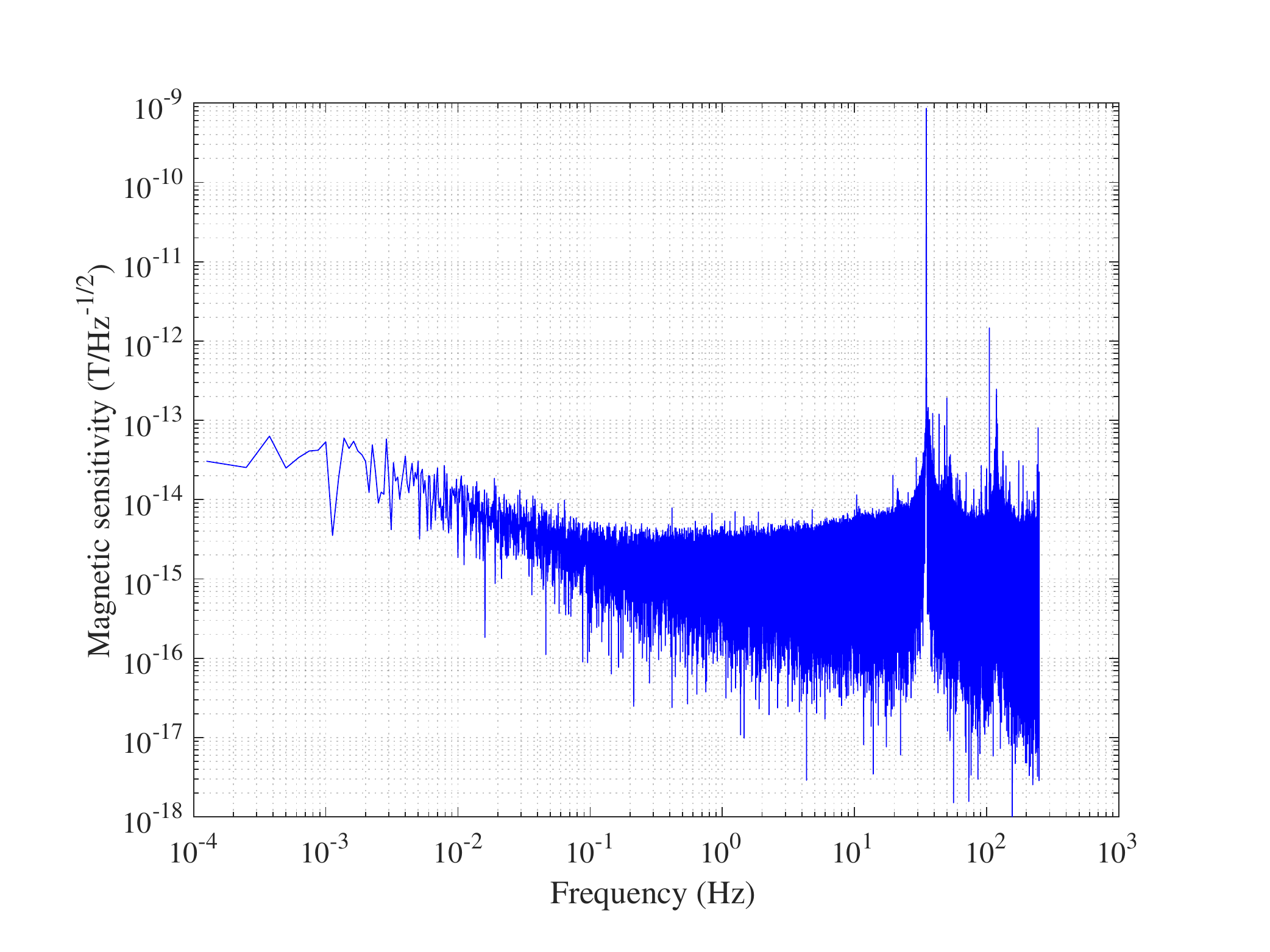}
\caption{(color online) The Rb-Xe comagnetometer sensitivity measured as the power spectral density of noise equivalent magnetic field.  The magnetic field sensitivity reached $\le$10 fT/Hz$^{1/2}$ at a frequency range from 0.01 to 10 Hz, important for various applications from biology to fundamental physics. Note the spike at $\sim$35 Hz is the $^{129}$Xe spin resonance peak.}
\label{fig4}
\end{figure}

With standard Fourier analysis and data fitting process, we extract the spin oscillation frequency versus phase and find it agree with the simulation results in Fig.\ref{fig1}(b). As shown in Fig.\ref{fig3}(c), a spectral linewidth of 0.04 Hz with SNR$\sim$1200 and a spectral linewidth of 0.53 mHz with SNR$\sim$40600 are obtained for the open loop and close loop signals, respectively. Compared to the open loop operation, the frequency resolution of the $^{129}$Xe spin resonance frequency was improved by a factor of $\sim$2540, from 33.3 $\mu$Hz down to 13.1 nHz, approaching the state of the art accuracy \cite{Limes2018PRL}. For $^{129}$Xe with gyromagnetic ratio 11.78 MHz/T, this level of frequency resolution leads to a magnetic field resolution of $\approx$1.11 fT, sufficient for applications such as the detection of human brain magnetic field \cite{MEG2006APL}. The noise equivalent magnetic field power spectral density is given in Fig.\ref{fig4}, showing a magnetic sensitivity of $\le$10 fT/Hz$^{1/2}$ acheived at the frequency range from 0.01 to 10 Hz, which is helpful for various applications from biomagnetism detection to fundamental physics.

It shall be noted that the center frequency of close-loop and open-loop signals does not coincide exactly with each other due to the limited tuning resolution of phase control by rheostats in present experiment. This shall be easily improved with higher phase tuning techniques, such as a direct digital synthesis (DDS). 

We have been observing the spin oscillation signal with a real-time oscilloscope for hours and found no sign of decaying at all, which means that the effective coherence time of spin oscillation probably is infinite. Once the loop is open, the spin oscillation starts to decay exponentially again within the intrinsic coherence time of noble gas spin. In this sense, the self-driving comagnetometer can be taken as a hybrid atomic spin oscillator, preferably working in continuous wave mode like conventional laser or maser. 

\begin{figure}[htbp]
\centering
\includegraphics[width=0.5\textwidth]{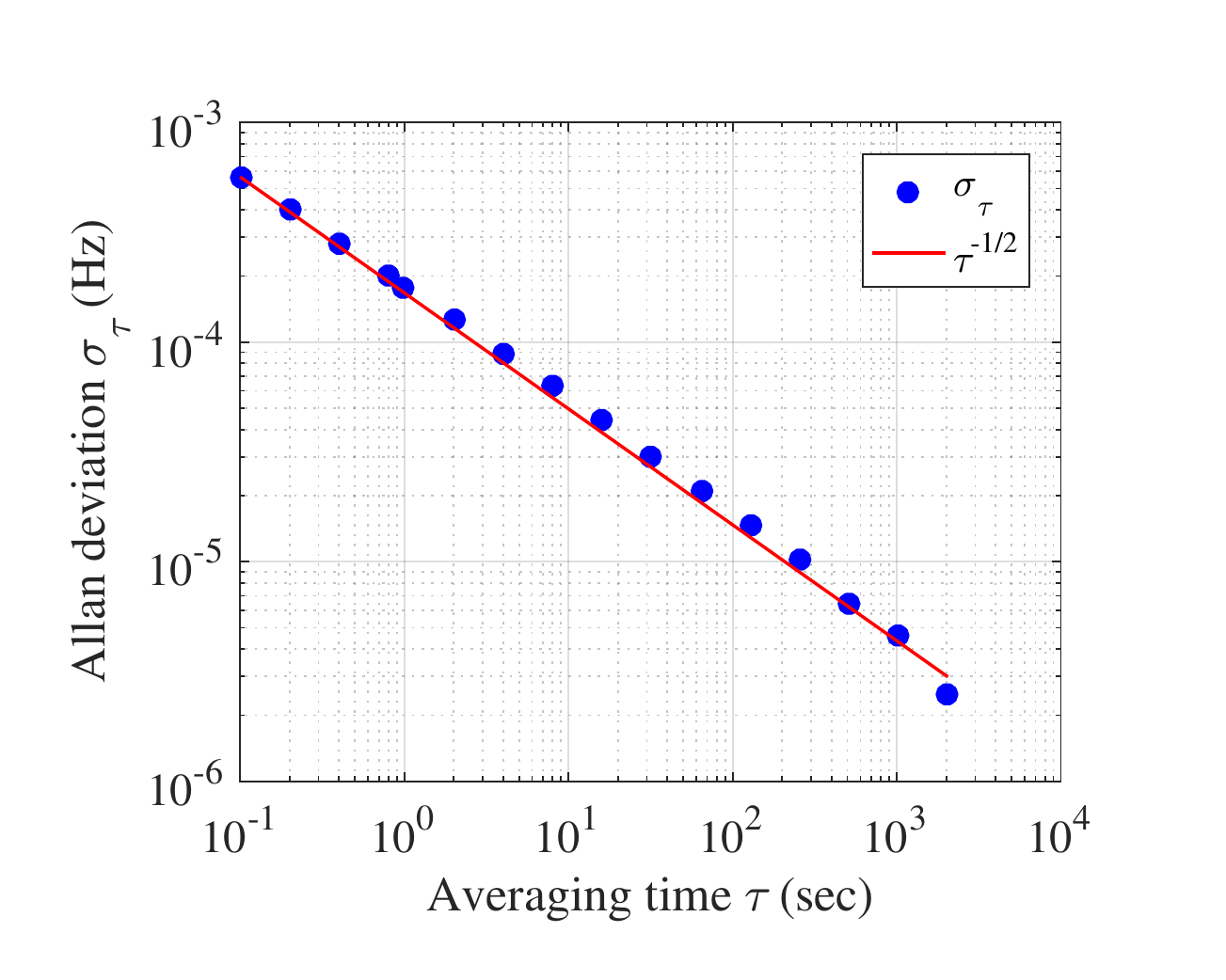}
\caption{(color online) Allan deviation of the $^{129}$Xe spin oscillation frequency in averaging time up to 1000 seconds. A frequency instability of 2.99 $\mu$Hz is achieved at a measuring time of 2048 seconds, equivalent to a bias instability of 3.87 $^{\circ}$/h.}
\label{fig5}
\end{figure}

To test the performance of hybrid spin oscillator in long term operation, we recorded continuously the spin oscillation for 10000 seconds in the close loop operation and excuted a standard Allan deviation analysis for the last 9000 seconds, as shown in Fig.\ref{fig5}. The frequency instability of spin oscillation reaches 2.99 $\mu$Hz at 2048 seconds averaging time, equivalent to a bias instability of $\sim$3.87 $^{\rm{o}}$/h for gyroscopic measurement.

With respect to the 13.1 nHz frequency resolution, the 2.99 $\mu$Hz frequency instability is relatively high. We attributed this deterioration to various frequency drift sources. For example, we observed for two days a correlation between the drift of spin oscillation frequency (at 10$^{-4}$ level) and the drift of heater power for vapor cell, which was a result of the slowly varying residual magnetic field produced by the leaky current of the heater wires. Besides, the pump laser power is in free running mode, whose fluctuation can cause significant fluctuation (at 1$\%$ level) of the optical pumping rate thus the fluctuation of alkali atomic spin magnetization, which finally cause the frequency shift of the noble gas spin oscillation. There are other factors affecting the middle to long term frequency instability, such as the drift of current feeding the bias field coils. 

As indicated by Fig.\ref{fig1}(a), the fluctuation of unlocked phase can also lead to a frequency shift with a slope of $\delta\nu$/$\delta\theta$$\sim$10$^{-4}$ Hz/$^{\circ}$. A possible improvement is to use phase-lock method \cite{Zhao2020SR} to lock the driving phase at the ZFS point. Considering the infinite effective coherence time, the self-driving spin oscillator has the potential to reach a frequency instability at nHz level at long term running. As shown by Fig.\ref{fig4}, the $\sigma_{\tau}$ continues going down at 2048 seconds, indicating a better frequency instability can be reached given longer running time. 

During the process of preparing the manuscript, we noticed the work by Jiang et al.\cite{Floquetmaser2021}, which presented similar phenomena. We have to emphasize the following important differences between their work and ours. First, we built the spin oscillator with a simpler prototype setup, neither external driving field nor parametric modulation (together with lock-in detection) is used in experiment and no significant SNR loss observed. Secondly, we derived the self-oscillating conditions for Rb-Xe comagnetometer by developing a different theoretical framework, especially the introduction of $G$ and $\theta$ parameters, which are conceptually convenient for understanding and easier to be used for guiding experiment. At last, we found the existence of zero frequency shift phase in theory and experiment, which is important for various applications in precision measurement physics.

Thanks for referee's kind reminding on the similar works on by Chupp et al.\cite{Zeemanmaser1994} and by Sato et al.\cite{spinmaser2018}. We would like to discuss the differences and compare with them generally in the following points.

Compared to the $^3$He and $^{129}$Xe Zeeman masers in \cite{Zeemanmaser1994}, the most important differences between their and our schemes are the detection methods. We have Rb spins and Xe spins in a single vapor cell, with Rb spins as the probe for Xe spin precession signals. We also used optical polarimeter for reading out the spin precession signal, instead of the traditional pick-up coils based on Faraday's electromagnetic induction \cite{Zeemanmaser1994}. Both arrangements shall enhance the detection sensitivity theoretically. However, on the other hand, the single cell schem also doesn't allow us to optimize the pump and probe parameters separately, which could be done with the separate pump and probe cells scheme in \cite{Zeemanmaser1994}. Single cell arrangement is also disadvantageous due to the back-action of Rb spin magnetization on the Xe spin precession in detection period. Besides, we used single nuclear spin specie instead of two nuclear spin species. This give our scheme more sensitivity to the magnetic field drift. Both arrangements can worsen the long term frequency stability in our scheme.

Compared to the colocated $^{129}$Xe and $^{131}$Xe spin masers by Sato et al.\cite{spinmaser2018}, the major difference is their introduce of the $^{131}$Xe spins in addtion to the $^{129}$Xe spins. This scheme theoretically can decrease the systematic frequency errors due to the magnetic field fluctuation, however, the nonzero nuclear quadrupole moment of $^{131}$Xe can render the sytem additional sensitivity to stray electric fields, worsening the long term frequency stability. Besides, they used a transmission detection on the probe laser, which could cause more noise compared to the polarimeter detection in our scheme. Costly photo-elastic modulation and lock-in detection was also used to improve the signal to noise ration in \cite{spinmaser2018}, while our scheme doesn't use any external modulation or lock-in detection and still reaches the same level of signal output of $\sim$1 Volt. 

Generally, the demonstrated self-driven hybrid atomic spin oscillator is still in its preliminary status, needing further optimizations in some details including the magnetic field drift control and laser power stabilization. However, it has already shown great potential in precision measurement with a simple-to-establish setup. In principle, the demonstrated spin oscillator scheme can be also applied to other dual-spin system, such as the K-$^3$He comagnetometer, and also to the trispin comagnetometer, such as the Rb-$^3$He-$^{129}$Xe or Rb-$^{129}$Xe-$^{131}$Xe configuration \cite{Limes2018PRL, Sheng2021PRA} with careful design of dual-channel driving electronics with respect to the two noble gas spin oscillation frequencies. 

In conclusion, we have demonstrated theoretically and experimentally a self-driving spin oscillator based on the Rb-Xe comagnetometer. A self-sustaining oscillator with infinite coherence time was realized at proper driving conditions. The spin oscillation frequency shift vanishes at certain phase points despite the Bloch-Siegert shift effect. The frequency resolution of the hybrid spin oscillator reaches $\sim$10 nHz level, potentially enhancing the detection sensitivity for magnetic or gyroscopic measurements with a simple apparatus. The magnetic sensitivity reaches $\le$10 fT/Hz$^{1/2}$ level at the frequency range from 0.01 to 10 Hz, very useful for various applications from biomagnetism detection to fundamental physics. With further improvement on the frequency instability, the hybrid atomic spin oscillator can work like laser or maser for long time operation, promising for various scientific or practical applications, such as the searching for new spin-dependent interactions and earth rotation monitoring.

The author Guobin Liu would like to thank Dong Sheng and Min Jiang from the University of Science and Technology of China for helpful discussions. The authors also appreciate the financial support by the Chinese Academy of Sciences under grant no E209YC1101 and by the National Time Service Center under grant no E024DK1S01.

%

\end{document}